\documentclass[letterpaper,notitlepage,nofootinbib,preprintnumbers,superscriptaddress]{revtex4-1} 

\usepackage[utf8]{inputenc}
\usepackage{amsmath,amssymb,amsfonts}

\usepackage{graphicx}

\usepackage{dcolumn}
\usepackage{bm}
\usepackage{color}
\usepackage{bbold}
\usepackage{url}

\usepackage{slashed}
\usepackage[svgnames]{xcolor}
\usepackage[english]{babel}
\usepackage{microtype}

\usepackage[colorlinks=true,linkcolor=blue,urlcolor=blue,citecolor=blue]{hyperref}

\begin{document}

\title{\boldmath $D$-term, strong forces in the nucleon, and their applications}

\author{Maxim V.~Polyakov}
	\affiliation{Petersburg Nuclear Physics Institute, 
		Gatchina, 188300, St.~Petersburg, Russia}
	\affiliation{Institut f\"ur Theoretische Physik II, 
		Ruhr-Universit\"at Bochum, D-44780 Bochum, Germany}
\author{Peter~Schweitzer}
	\affiliation{Department of Physics, University of Connecticut, 
		Storrs, CT 06269, USA}

\begin{abstract}
The $D$-term is a fundamental particle property which is defined 
through the matrix elements of the energy-momentum tensor and as
such in principle on equal footing with mass and spin. 
Yet the experimental information on the $D$-term of any hadron is 
very scarce. The $D$-term of the nucleon 
can be inferred from studies of hard-exclusive reactions, and its
measurement will give valuable insights on the dynamics, structure,
and the internal forces inside the nucleon. We review the latest 
developments and the fascinating applications of the $D$-term and 
other energy-momentum tensor (EMT) form factors. We also suggest
a definition of the mechanical mean square radius and make a 
prediction for its size.
\end{abstract}

\date{December 2017} 

\maketitle

%===============================================================================
%===============================================================================

\section{Introduction}

The matrix elements of the energy-momentum tensor (EMT) \cite{Pagels:1966zza}
provide the most fundamental information: namely the mass and spin of a
particle. The EMT matrix elements contain, for spin 0 and $\frac12$,
one more fundamental information: the $D$-term  \cite{Polyakov:1999gs}, 
which is not known experimentally for any particle. The EMT form factors
of the nucleon are given by
(it is $P=\frac12(p+p')$, $\Delta=p'-p$, $t=\Delta^2$, $\bar u u =2 m$,
$a_{\{\mu} b_{\nu\}}=a_\mu b_\nu + a_\nu b_\mu$, see \cite{Hudson:2016gnq} 
for a review and other notations)
\begin{align}
    \langle p^\prime| \hat T_{\mu\nu}^{q,g}(0) |p\rangle
    = \bar u{ }^\prime\biggl[
      A^{q,g}(t)\,\frac{\gamma_{\{\mu} P_{\nu\}}}{2}
    + B^{q,g}(t)\,\frac{i\,P_{\{\mu}\sigma_{\nu\}\rho}\Delta^\rho}{4m}
    + D^{q,g}(t)\,\frac{\Delta_\mu\Delta_\nu-g_{\mu\nu}\Delta^2}{4m}
    + {\bar c}^{q,g}(t)\,g_{\mu\nu} \biggr]u\,.
    \label{Eq:EMT-FFs-def}
\end{align}
The total form factors $A(t)\equiv \sum_a A^a(t)$ with 
$a=g,\,u,\,d,\,\dots\,$, etc are renormalization scale independent. 
The constraints $A(0)=1$ and $B(0)=0$ (vanishing of 
"anomalous gravitomagnetic moment") reflect that the spin of 
the nucleon is $\frac12$ and its energy is $m$ in the rest frame.
The $D$-term $D\equiv D(0)$ is unconstrained, and must be determined 
from experiment. Only the total EMT is conserved, and the form factors 
$\bar{c}^a(t)$ satisfy $\sum_a\bar{c}^a(t)=0$.

The most natural way to probe EMT form factors, scattering off gravitons, 
is also the least practical, see Fig.~\ref{Fig-PS-1}a. An opportunity 
to access EMT form factors emerged with the advent of GPDs 
\cite{Mueller:1998fv,Ji:1996ek,Radyushkin:1996nd,Radyushkin:1996ru,
Ji:1996nm,Collins:1996fb,Radyushkin:1997ki}
which describe hard-exclusive reactions, Fig.~\ref{Fig-PS-1}b.
The second Mellin moments of unpolarized quark GPDs yield the EMT form factors 
(gluons analog),
\begin{align}
	\int{\rm d}x\;x\, H^q(x,\xi,t) = A^q(t) + \xi^2 D^q(t) \,, \quad \quad
        \int{\rm d}x\;x\, E^q(x,\xi,t) = B^q(t) - \xi^2 D^q(t) \,.
    	\label{Eq:GPD-Mellin}
\end{align}
Adding up the two equations in (\ref{Eq:GPD-Mellin}) and extrapolating
$t\to 0$ provides the key to the nucleon's spin decomposition 
\cite{Ji:1996ek}. But what does the $D$-term mean? In the 
next section we review what is known
.

\section{The $\mathbf{D}$-term}

The $D$-term is a contribution to unpolarized GPDs in the region 
$-\xi \le x \le \xi$ \cite{Polyakov:1999gs}, and determines their asymptotics 
in the limit of renormalization scale $\mu\to\infty$ \cite{Goeke:2001tz}.
It appears in Radon transforms \cite{Teryaev:2001qm}, and encodes the 
mechanical properties of a particle \cite{Polyakov:2002yz} 
as we will discuss below in detail. 
It is a subtraction constant in fixed-$t$ dispersion relations 
for DVCS amplitudes \cite{Anikin:2007yh,Diehl:2007jb,Radyushkin:2011dh},
and related to fixed poles in the angular momentum plane in 
virtual Compton scattering discussed in 
\cite{Cheng:1970vg,Brodsky:1971zh,Brodsky:1972vv},
however, it was shown that the $J = 0$ fixed pole universality 
hypothesis is an external assumption and might never be proven 
theoretically \cite{Muller:2015vha}.

First principle information on the $D$-term of the nucleon is 
limited to the prediction from the QCD multi-color limit $N_c\to\infty$ 
of the flavor hierarchy $D^u(t)+D^d(t) \sim N_c^2$ being much larger than 
$D^u(t)-D^d(t) \sim N_c$ \cite{Goeke:2001tz}, and to results on $D^q(t)$ 
from lattice QCD \cite{Hagler:2003jd,Gockeler:2003jfa} and dispersion 
relations \cite{Pasquini:2014vua}. 

The $D$-term is of importance 
for the phenomenological description of hard-exclusive reactions 
\cite{Ellinghaus:2002bq,Kumericki:2007sa,Guidal:2013rya},
but cannot yet be extracted in model-independent way.

\newpage
\section{Mechanical properties}

Mechanical properties are studied by Fourier transforming 
(\ref{Eq:EMT-FFs-def}) in Breit frame (where $\Delta^0=0$)
with respect to~$\vec{\Delta}$ which yields the static EMT 
whose $ij$-components define the stress tensor \cite{Polyakov:2002yz}.
The total quark + gluon stress tensor can be decomposed in a 
traceless part associated with shear forces $s(r)$ and a trace 
associated with the pressure $p(r)$,
\begin{align}
	T^{ij}(\vec{r}) = \biggl(%e^ie^j
	\frac{r^ir^j}{r^2}-\frac13\,\delta^{ij}\biggr) s(r)
	+ \delta^{ij}\,p(r)\,.
    	\label{Eq:stress-tensor-p-s}
\end{align}
The shear forces are ``good observables'' and 
exist also separately for quarks and gluons (although 
$T_{\mu\nu}^q$ and $T_{\mu\nu}^g$ are not conserved separately,
the EMT-nonconserving $g_{\mu\nu}\,\bar{c}^a(t)$ in
(\ref{Eq:EMT-FFs-def}) drop out from the traceless part of the 
stress tensor). The shear forces also allow one to define the quark 
and gluon contributions to the $D$-term  \cite{Polyakov:2002yz},
\begin{align}
	D^{q,g} = - \frac{2}{5}\,m \int{\rm d}^3r\;T^{q,g}_{ij}(\vec{r})\,
	\biggl(r^ir^j-\frac13\,r^2\delta^{ij}\biggr) \,.
    	\label{Eq:stress-tensor-D-term}
\end{align}
In contrast to this $p(r)$ is defined only for the total system, and 
has no relation to the separate $D^q$ and $D^g$ \cite{Polyakov:2002yz}.	
EMT conservation, $\partial^\mu\hat{T}_{\mu\nu}=0$, implies 
the relation $\frac23\,s^\prime(r)+\frac2r\,s(r)+p^\prime(r)=0$ and the
von Laue condition $\int_0^\infty {\rm d}r\,r^2p(r)=0$, a necessary 
condition for stability, which shows how the internal forces balance.
The 3D density interpretation is subject to relativistic corrections
which are acceptably small for the nucleon and nuclei \cite{Hudson:2017xug}.

\section{The mechanical mean square radius}

The normal component of the total force exhibited by the system on 
an infinitesimal piece of area  $dS$ at the distance~$r$ has the form  
$F^i(\vec r)=T^{ij}(\vec r)\;r^j/r\,dS=\left[\frac23s(r)+p(r)\right]\,r^i/r\,dS$.
In Ref.~\cite{Perevalova:2016dln} we argued that for the mechanical stability 
of the system the corresponding force must be directed outwards. Therefore
the local criterion for the mechanical stability can be formulated as the 
inequality $\frac 23 s(r) +p(r) >0$ .
This inequality implies that the $D$-term for any stable system must be 
negative \cite{Perevalova:2016dln}. Additionally, the positive combination
$\left[\frac 23 s(r) +p(r)\right]$ has the meaning of the force distribution 
in the system. That allow us to introduce the notion of the mechanical
radius for hadrons:
\begin{equation}
\label{eq:mechanicalradius}
	\langle r^2\rangle_{\rm mech} 
	=  \frac{\int d^3r\  r^2\ \left[\frac 23 s(r) +p(r)\right]}
		{\int d^3r\ \left[\frac 23 s(r) +p(r)\right]}
	= - \,\frac 94 \;\frac{D}{m\int d^3r\ s(r)} =\frac{6 D}{\int_{-\infty}^0 dt\ D(t)} \, ,
\end{equation}
where in the last equality we used the relation of the surface tension 
energy of the system $\int d^3r\ s(r)$ to $D(t)$ (for definition see 
Eq.~(\ref{Eq:EMT-FFs-def})):
\begin{equation}
\label{eq:surfacetensionEnergy}
	\int d^3r\ s(r)=-\frac{3}{8 m} \int_{-\infty}^0 dt\ D(t)	 \, ,
\end{equation}
We see that $D$-term also determines the mechanical 
radius of the hadrons. Notice the unusual definition: 
unlike e.g.\ the mean square charge radius, the mechanical 
mean square radius is not related to the slope of a form factor.

The definition (\ref{eq:mechanicalradius}) of the 
mechanical radius applied to liquid drop model for a nucleus gives
intuitively clear result $\langle r^2\rangle_{\rm mech}=\frac 35 R_{\rm drop}^2$.
The chiral quark soliton model predicts the mechanical radius of the 
proton to be about $25\,\%$ smaller than its mean square charge radius: 
$\langle r^2\rangle_{\rm mech} \approx 0.75 \, \langle r^2\rangle_{\rm charge}$.

In Ref.~\cite{Kumano:2017lhr}, where an extraction of EMT form 
factors of $\pi^0$ in the time-like region was reported, also a definition
of a ``mechanical radius'' was proposed, however, in terms of the slope 
of the form factor $D(t)$. This differs from our definition 
(\ref{eq:mechanicalradius}), and is not an appropriate measure 
of the true mechanical radius of a hadron \cite{NEW}.

\section{Distinguishing bosons and fermions}

Before discussing the forces in nuclei and nucleons it is
instructive to inspect free field theories. Interestingly, the 
particle property $D$-term can ``distinguish'' between 
elementary pointlike bosons and fermions in the following sense. 
A free spinless boson has a non-zero intrinsic $D$-term $D=-1$.
In sharp contrast to this, the $D$-term of a free spin-$\frac12$ 
fermion is zero, see \cite{Hudson:2017xug,Hudson:2017oul} and 
references therein.

This unexpected finding deserves a comment. One can give a pointlike 
boson a ``finite, extended, internal structure'' (by ``smearing out'' 
its energy density $T_{00}(r)=m\,\delta^{(3)}(\vec{r})$ with e.g.\ a 
narrow Gaussian, and analog for other densities) such that the property 
$D=-1$ is preserved. This yields automatically the characteristic shapes 
for pressure and other densities found in dynamical model calculations 
\cite{Hudson:2017xug}.
Field theories of extended (solitonic $Q$-ball type) solutions 
with $D=-1$ can be constructed where such ``smearing out'' is
implemented \cite{Hudson:2017xug}.
In general though, interactions modify the free boson value 
$D=-1$ \cite{Mai:2012yc,Mai:2012cx,Cantara:2015sna}.
But the situation is fundamentally distinct for fermions: here 
interactions do not modify the $D$-term, they {\it generate} it. 
A non-zero fermionic $D$-term is of dynamical origin 
\cite{Hudson:2017oul}. Recalling that all known matter is fermionic,
this indicates the importance to study the physics of the $D$-term.

\section{Strong forces in nuclei and nucleon}

The liquid drop model of a large nucleus (atomic number $A$, 
mass $m\sim A$, radius $R\sim A^{1/3}$, surface tension $\gamma$ 
from Bethe-Weizs\"acker formula) illustrates how interactions 
generate the $D$-term: the shear forces are $s(r)=\gamma\,\delta(r-R)$ and
$D\equiv-\frac{4}{15}\,m\int{\rm d}^3r\,r^2s(r)=-\frac{5\pi}{3}\,m\,\gamma\,R^4$
grows as $D\sim A^{7/3}$ \cite{Polyakov:2002yz} which more
sophisticated nuclear models confirm \cite{Guzey:2005ba}.
The $D$-term of the nucleon was studied in the bag,
chiral quark soliton, Skyrme model
\cite{Ji:1997gm,Petrov:1998kf,Schweitzer:2002nm,Ossmann:2004bp,
Goeke:2007fp,Goeke:2007fq,Wakamatsu:2007uc,Cebulla:2007ei,Jung:2013bya},
and nuclear medium modifications were investigated 
\cite{Kim:2012ts,Jung:2014jja}. 
The interactions in these models could not be more diverse,
but in all cases $D$ was found negative. Also the $D$-terms
of the photon and the $\Delta$-resonance are negative
\cite{Gabdrakhmanov:2012aa,Perevalova:2016dln}.
Chiral perturbation theory predicts the $D$-terms of Goldstone 
bosons \cite{Novikov:1980fa,Voloshin:1982eb,Donoghue:1991qv} but not 
of other hadrons \cite{Belitsky:2002jp,Ando:2006sk,Diehl:2006ya}.

%-------------------------------------------------------------------------------
\begin{figure}[t!]
\centering
\includegraphics[height=3cm]{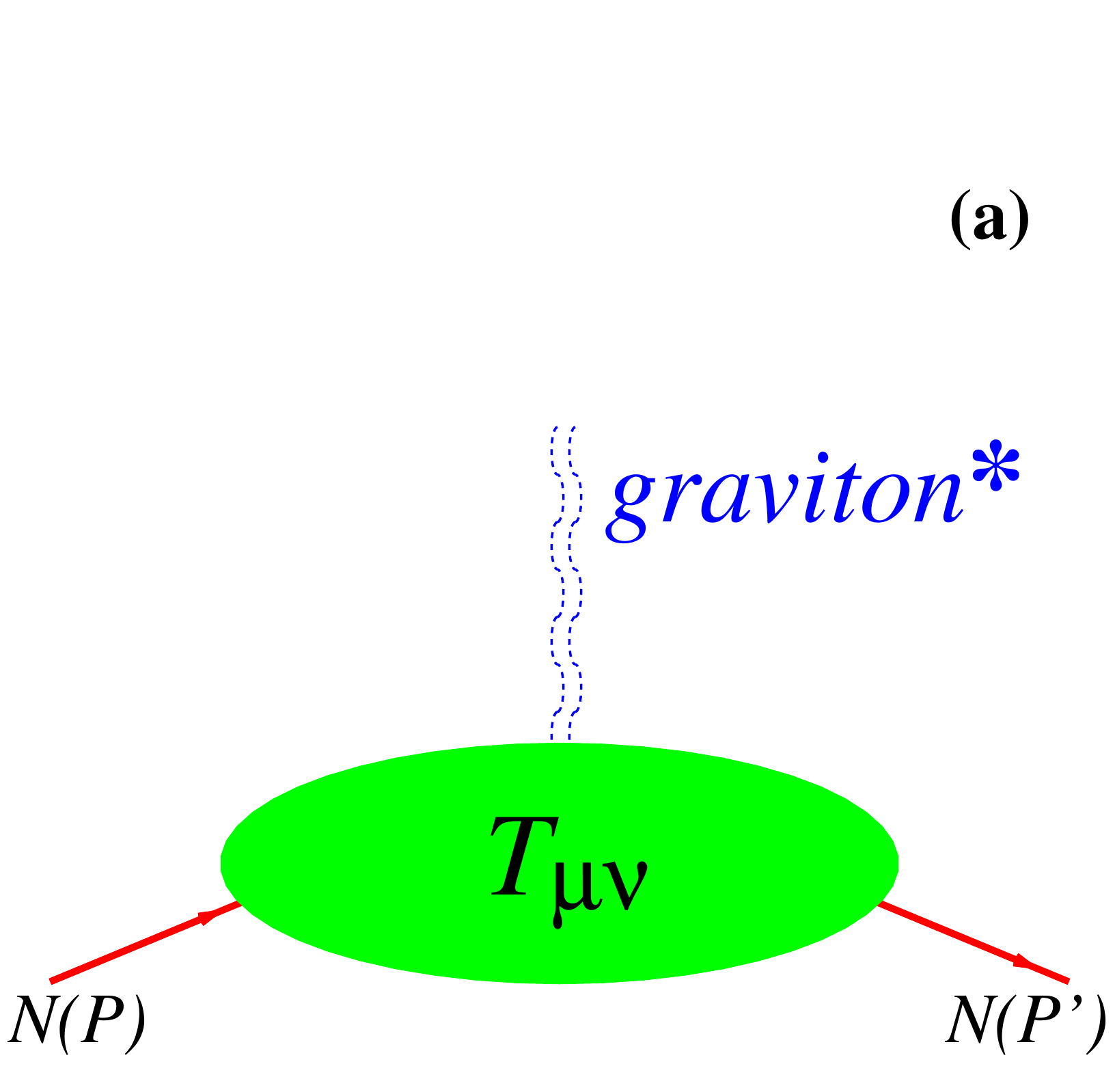} \ \ \ \
\includegraphics[height=3cm]{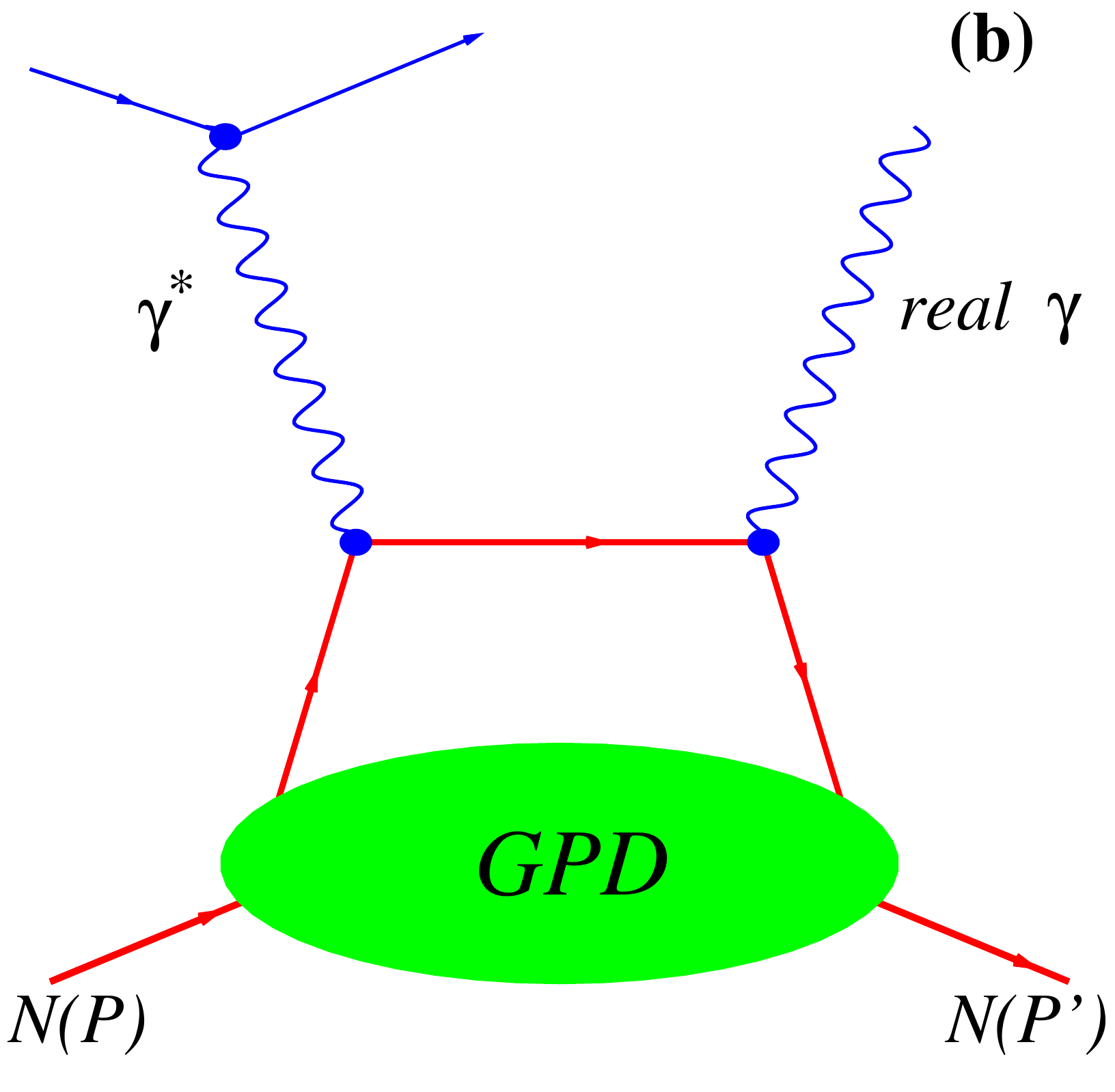} \ \ \ \
\includegraphics[height=3cm]{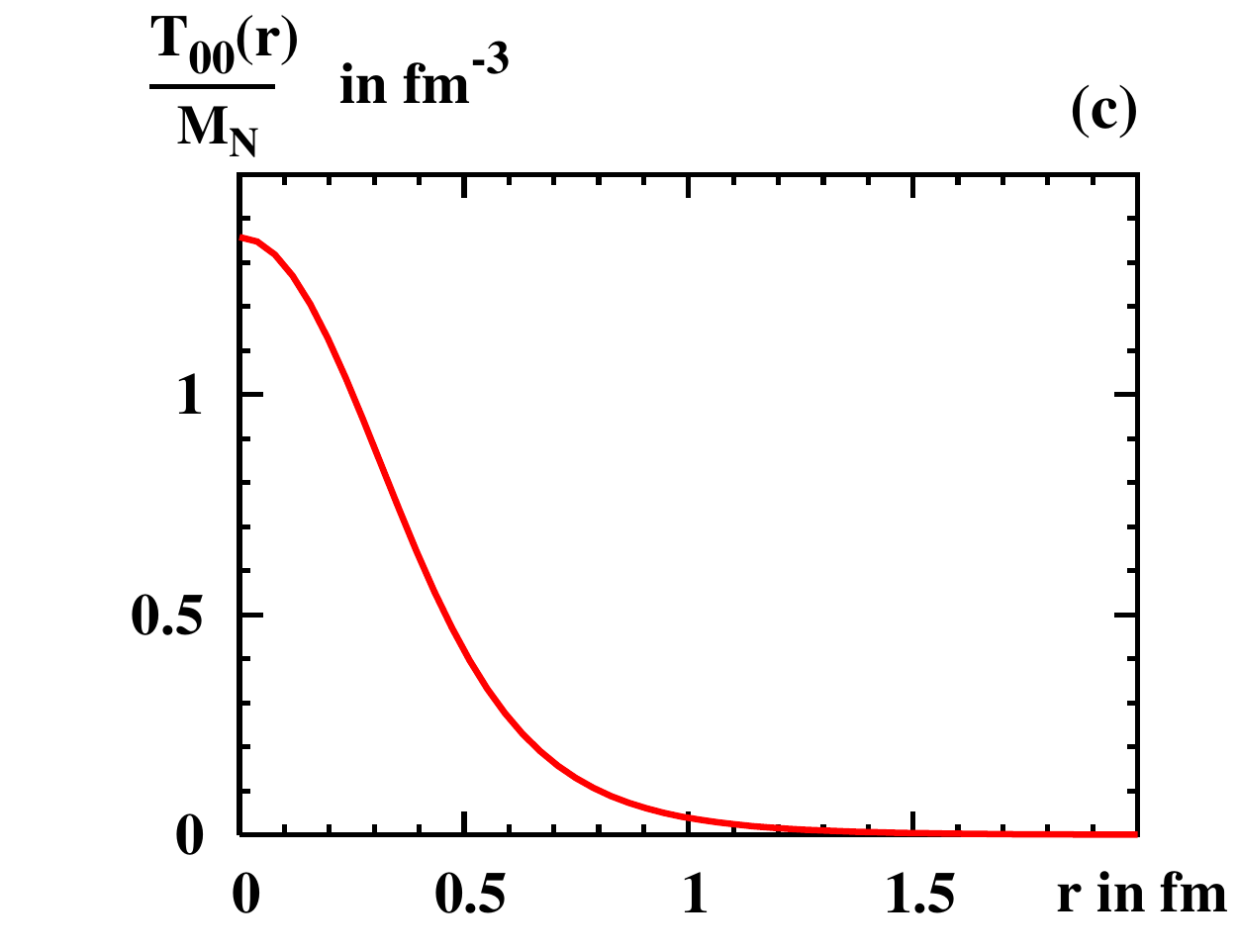} \ \ \ \
\includegraphics[height=3cm]{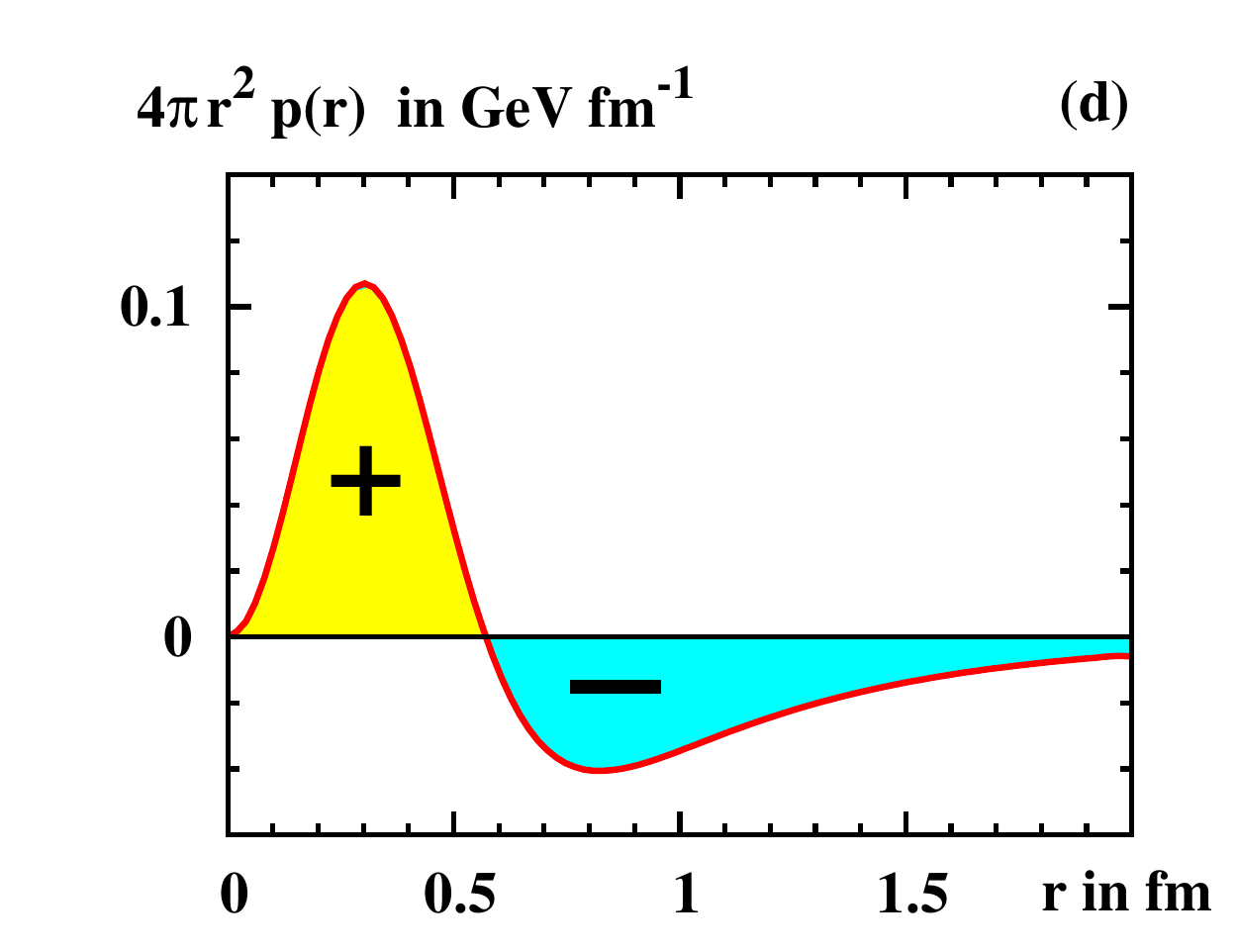}
\caption{(Colour online) 
	A natural but impractical probe of EMT form factors is 
	scattering off gravitons (a). Hard-exclusive reactions 
	like deeply virtual Compton scattering (DVCS) provide 
	a realistic way to access EMT form factors (b).
	Energy density (c) and pressure distribution (d) in 
	the nucleon according to calculations in the chiral 
	quark soliton model \cite{Goeke:2007fp}.}
\label{Fig-PS-1}
\end{figure}
%-------------------------------------------------------------------------------

Figs.~\ref{Fig-PS-1}c and \ref{Fig-PS-1}d show $T_{00}(r)$
and $p(r)$ from the chiral quark soliton model \cite{Goeke:2007fp}. 
In the center $T_{00}(0) = 1.7\,{\rm GeV/fm}^3$ which is around
13 times the nuclear matter density while $p(0) = 0.23 {\rm GeV/fm}^3$.
The pressure is positive in the center of the nucleon which means
repulsion, and negative for $r \gtrsim 0.6\,{\rm fm}$ 
which means attraction. Repulsive and attractive forces 
balance each other exactly according to the von 
Laue condition $\int_0^\infty {\rm d}r\,r^2p(r)=0$.
Notice that the ``hydrostatic pressure force'' 
$4\pi r^2p(r)$ reaches at most about $0.1\,{\rm GeV/fm}$
to be compared to the QCD string tension $k\sim 1\,{\rm GeV/fm}$, 
i.e.\ in a ground state hadron like nucleon only a fraction of the 
strong confining forces is needed to achieve equilibrium.

\section{EMT densities and applications to hidden-charm pentaquarks}

The extraction of the $D$-term will not only provide insights on
how internal strong forces balance inside the~nucleon. Knowledge 
of EMT form factors has also applications to the spectroscopy of the 
exotic hidden-charm pentaquarks observed by LHCb \cite{Aaij:2015tga} 
which decay in $J/\psi$ and proton. One of these states, 
the narrow $P^+_c(4450)$ with $\Gamma\sim 40\,{\rm MeV}$, can be 
described exploring that quarkonia are small compared to the nucleon 
size justifying a multipole expansion which shows that the 
baryon-quarkonium interaction is dominated by the emission of two 
virtual chromoelectric dipole gluons in a color singlet state. 
The effective interaction \cite{Voloshin:1979uv} can be 
expressed in terms of the quarkonium chromoelectric polarizability 
$\alpha$ and nucleon EMT densities as 
$V_{\rm eff} = -\,\alpha\,\frac{4\pi^2}{b}\,(\frac{g}{g_s})^2
(\nu\,T_{00}(r)-3\,p(r))$. Here $b = \frac{11}{3}N_c-\frac23 N_f$ 
is the leading coefficient of the Gell-Mann-Low function, 
$g_s$ ($g$) is the strong coupling constant at the scale 
associated with the nucleon ($J/\Psi$) size, and the parameter 
$\nu$ was estimated $\nu\sim 1.5$ \cite{Eides:2015dtr}.

For realistic values of $\alpha(1S)$, $V_{\rm eff}$ is not strong 
enough to bind the nucleon and $J/\Psi$ with the EMT densities 
from the chiral quark soliton. But a bound state of the mass
$4450\,{\rm MeV}$ exists in the $\psi(2S)$-nucleon channel for 
$\alpha(2S)\sim 17\,{\rm GeV}^{-3}$ \cite{Eides:2015dtr} which is 
close to perturbative QCD estimates of this chromoelectric 
polarizability \cite{Peskin:1979va,Bhanot:1979vb}. 
The decay of $P^+_c(4450)$
is governed by the same $V_{\rm eff}$ but with a much smaller
transition polarizability $\alpha(2S\to1S)\sim {\cal O}(1)$ 
\cite{Peskin:1979va,Bhanot:1979vb,Voloshin:2007dx} which
explains the relatively narrow width. The other putative
pentaquark state seen by LHCb, $P^+_c(4380)$, is much broader
with $\Gamma\sim 200\,{\rm MeV}$ and not described by this
binding mechanism \cite{Eides:2015dtr}. 
These findings are confirmed in the Skyrme model indicating 
they are largely model-insensitive \cite{Perevalova:2016dln}.
The approach makes definite predictions for similar bound states 
of $\psi(2S)$ with $\Delta$ \cite{Perevalova:2016dln} and 
hyperons \cite{Eides:2017xnt} which will allow us to test 
the theoretical framework.

\section{Conclusions}

We have reviewed aspects of the physics associated with the $D$-term 
and other EMT properties. We have also given a definition of the
mechanical radius of a hadron. The $D$-term and the other EMT form
factors are  highly fascinating, and will provide exciting insights 
on the strong forces in the nucleon and nuclei from a so far unexplored
perspective. The physics associated with EMT form factors also
offers important applications.

\ \\
\noindent{\bf Acknowledgments.}
%\begin{acknowledgments}
The work reviewed here was partly supported by NSF, Contract No.\ 1406298 
and by CRC110 (DFG).
%\end{acknowledgments}

\end{document}